\documentclass[a4paper,12pt]{article}
\usepackage{amssymb}
\usepackage{graphicx}

\def\journal#1#2#3#4{{\it #1} {\bf #2}, #3 #4}

\def\grad{\mathbf{\nabla}}
\def\o{\mathbf{\omega}}
\def\v{\mathbf{v}}
\def\U{\mathbf{U}}

\def\E{\mathbf{E}}
\def\B{\mathbf{B}}
\def\F{\mathbf{F}}
\def\Q{\mathrm{Q}}
\def\x{\mathbf{x}}

\def\l{{\cal L}}
\def\H{{\cal H}}
\def\D{{\cal D}}
\def\Z{{\cal Z}}
\def\J{{\cal J}}
\def\sf{\mathrm{F}}
\def\sg{\mathrm{G}}
\def\sx{\mathrm{X}}
\def\Jb{\mathbf{\cal J}}
\def\pd{\partial}
\def\pds{{\partial\hspace{-2.2mm}/ \, }}
\def\d{\mathrm{d}}
\def\Jb{\mathbf{J}}
\def\exp{\mathrm{exp}}

\def\be{\begin{equation}}
\def\ee{\end{equation}}
\def\bea{\begin{eqnarray}}
\def\eea{\end{eqnarray}}
\def\ie{\textit{i.e.} }

\title{Magnetofluid Unification in the Yang-Mills Lagrangian}

\author{A. Fajarudin$^{a,b}$, 
A. Sulaiman$^{a,c,d}$\footnote{Email : lyman@tisda.org, sulaiman@teori.fisika.lipi.go.id}, 
T.P. Djun$^a$\footnote{Email : tpdjun@teori.fisika.lipi.go.id} \, and 
L.T. Handoko$^{a,b}$\footnote{Email : handoko@teori.fisika.lipi.go.id, laksana.tri.handoko@lipi.go.id}}
\date{}

\begin{document}

\renewcommand{\thesubsection}{\Roman{subsection}}
\maketitle

%\begin{picture}(0,0)
%       \put(310,180){FISIKALIPI-07017}
%\end{picture}

\thispagestyle{empty}

\begin{center}
\begin{small}
\noindent
$^{a)}$Group for Theoretical and Computational Physics, Research Center for Physics, Indonesian Institute of Sciences, Kompleks Puspiptek Serpong, Tangerang 15310, Indonesia\\
$^{b)}$Department of Physics, University of Indonesia, Kampus UI Depok, Depok 16424, Indonesia\\
$^{c)}$P3 TISDA BPPT, BPPT Bld. II (19$^{\rm th}$ floor), 
Jl. M.H. Thamrin 8, Jakarta 10340, Indonesia\\
$^{d)}$Department of Physics, Bandung Institute of Technology, Jl. Ganesha 10, Bandung 40132, Indonesia\\
\end{small}
\end{center}

\vspace*{5mm}

\begin{abstract}
The Yang-Mills magnetofluid unification is constructed using lagrangian approach by imposing certain gauge symmetry to the matter inside the fluid. The model provides a general description for relativistic fluid interacting with Abelian or non-Abelian gauge field. The differences with the hybrid magnetofluid model are discussed, and few physical consequences of this formalism are worked out.
\end{abstract}

\vspace*{5mm}

\vspace*{5mm}

Some experimental discoveries and studies suggest that the deconfined quark gluon matter is behaving more like a quark gluon plasma (QGP) liquid \cite{adler,adeox,shuryak}. This fact motivates tremendous works in constructing the non-Abelian fluid models like magnetohydrodynamics \cite{heinz, choquet, holm, blaizot, bistrovic, mahajan1, mahajan2, mahajan3, manuel}.

In some recent models \cite{mahajan1, mahajan2, mahajan3}, the relativistic hot fluid was described in terms of hybrid magnetofluid field which unifies the electromagnetic and fluid fields. The unification is represented by the effective field strength tensor, $M_{\mu\nu} \equiv F_{\mu\nu} + {m/q} S_{\mu\nu}$ combining appropriately weighted electromagnetic and fluid fields. The model has further been generalized to the non-Abelian case \cite{mahajan2}. Also its topological structures has been investigated in detail to model the glueballs \cite{mahajan3}.

In this Letter, we follow the same line but with different starting point using lagrangian approach and  imposing certain gauge symmetry on it. This approach is intended to improve the Yang-Mills magnetofluid unification model \cite{mahajan2} in two aspects :
\begin{itemize}
\item From field theory point of view, the fluid and another gauge fields should physically be different fields, and then should have independent kinetic terms in the lagrangian. Deploying the above effective field tensor would lead to the unphysical two point interaction connecting two different fields from the  gauge invariance kinetic term $M_{\mu\nu}M^{\mu\nu}$ in the lagrangian.
\item The hybrid strength tensor $M_{\mu\nu}$ suggests that the fluid and another Abelian or non-Abelian gauge fields are gauge invariance under a single gauge transformation and both should belong to the same Lie group. This requirement is actually necessary since the formalism relies on the Lorentz force equation. This also avoids us to deal with, for instance, the case of non-Abelian fluid interacting with an (Abelian) electromagnetic field.
\end{itemize}

In order to overcome such problems, we propose local and independent gauge symmetries for each field of fluid and another gauge boson interacting with it, inspired by the unified theory known in particle physics. The model is able to deal with some scenarios like : (i) non-Abelian fluid interacting with non-Abelian gauge field under G($n$)$_\mathrm{F}$ $\otimes$ G($n$)$_\mathrm{G}$ symmetry; (ii) non-Abelian fluid interacting with an (Abelian) electromagnetic field under G($n$)$_\mathrm{F}$ $\otimes$ U(1)$_\mathrm{G}$ symmetry; and (iii) Abelian fluid interacting with an electromagnetic field under U(1)$_\mathrm{F}$ $\otimes$ U(1)$_\mathrm{G}$ symmetry, \ie the original magnetofluid model. Here F and G denote the fluid and another gauge fields respectively. Note that G($n$) could be any $n$ dimensional Lie group like SU($n$).

Following the basic procedure in lagrangian approach, we start from the following matter lagrangians,
\be
        \l_\mathrm{matter} = \left\{
	\begin{array}{ll}
	{\displaystyle \left( \pd_\mu \Phi \right)^\dagger  \left( \pd^\mu \Phi \right) + \frac{1}{2} m_\Phi^2 \, \Phi^\dagger \Phi + V(\Phi)} & \mathrm{for \, boson} \\
	{\displaystyle i \overline{\Psi} \, \pds \Psi - m_{\Psi} \overline{\Psi} \Psi} & \mathrm{for \, fermion} \\
	\end{array}
	\right. \, .
        \label{eq:lmatter}
\ee
$V(\Phi)$ is the potential, for example in a typical $\Phi^4$ theory, $V(\Phi) = \frac{1}{4} \lambda (\Phi^\ast \Phi)^2$, and $\pds \equiv \gamma_\mu \pd^\mu$ with $\gamma_\mu$'s are the Dirac matrices.

Under a local non-Abelian G($n$) gauge transformation $U \equiv \exp[-i T_a \theta^a(x)] \approx 1 - i T_a\theta^a(x)$ with $\theta^a \ll 1$, the matter field is transformed as $\Phi \stackrel{U}{\longrightarrow} \Phi^\prime \equiv \exp[-i T_a \theta^a(x)] \, \Phi$ or $\Psi \stackrel{U}{\longrightarrow} \Psi^\prime \equiv \exp[-i T_a \theta^a(x)] \, \Psi$ with the matter field is in general an $n \times 1$ multiplet containing $n$ elements for $n$ dimension Lie groups as SU($n$), O($n+1$),  etc. $T^a$'s are generators belong to those Lie groups and satisfy certain commutation  relation $[T^a,T^b] = i f^{abc} T^c$ with $f^{abc}$ is the anti-symmetric structure constant. Anyway, the number of generators, and also gauge bosons, is determined by the dimension of group under consideration. For an SU($n$) group one has $n^2 - 1$ generators and the index $a$ runs over $1, 2, \cdots, n^2 - 1$. The U(1) is realized by a phase transformation, $T_a \theta^a(x) \rightarrow \theta(x)$. The gauge invariance is then revealed by introducing gauge field $A_\mu$'s which are transformed as $A^a_\mu \stackrel{U}{\longrightarrow} {A^a_\mu}^\prime \equiv A^a_\mu + (1/g) (\pd_\mu \theta^a) + f^{abc} \theta^b A^c_\mu$, and replacing the derivative with the covariant one, $\D_\mu \equiv \pd_\mu - i g \, T^a A^a_\mu$, where $g$ is the gauge ``charge''. Further the gauge invariance kinetic term for gauge boson $A^a_\mu$ takes the form of $F^a_{\mu\nu} {F^a}^{\mu\nu}$, with strength tensor $F^a_{\mu\nu} \equiv \pd_\mu A^a_\nu - \pd_\nu A^a_\mu + g f^{abc} A^b_\mu A^c_\nu$.

Now, let us consider the most general case of non-Abelian fluid interacting with any non-Abelian gauge fields represented by G($n$)$_\mathrm{F}$ $\otimes$ G($n$)$_\mathrm{G}$ symmetry. Hence we can deploy the above gauge principle for each field independently, that is the gauge transformations are performed separately with independent phase parameters $\theta(x)$'s. Imposing the symmetries, and assigning $U_\mu$ and $A_\mu$ as the gauge bosons associated with fluid and another gauge fields belong to different group spaces, the covariant derivative becomes,
\be
	\D_\mu \equiv \pd_\mu + i g_\sf \, T_\sf^a U^a_\mu + i g_\sg \, T_\sg^a A^a_\mu \, ,
\ee
where $g_\sf$ is the ``charge'' for fluid, while $g_\sg$ is the gauge charge.

Finally the total lagrangian with such gauge symmetries becomes,
\be
	\l = \l_\mathrm{matter} + \l_\mathrm{gauge} + \l_\mathrm{int.} \; ,
	\label{eq:l}
\ee
where, 
\bea
        \l_\mathrm{gauge} & = & - \frac{1}{4} S^a_{\mu\nu} {S^a}^{\mu\nu} 
				- \frac{1}{4} F^a_{\mu\nu} {F^a}^{\mu\nu} \; ,
        \label{eq:lfce} \\
	\l_\mathrm{int.} & = & - g_\sf {J^a_\sf}_\mu {U^a}^\mu 
				- g_\sg {J^a_\sg}_\mu {A^a}^\mu + \l_\mathrm{int}^\mathrm{boson} \; ,
        \label{eq:li}
\eea
and $S^a_{\mu\nu} \equiv \pd_\mu U^a_\nu - \pd_\nu U^a_\mu + g_\sf f^{abc} U^b_\mu U^c_\nu$. Meanwhile, in the case of bosonic matter there are additional interference terms in Eq. (\ref{eq:li}) coming from the gauge invariance kinetic term $(\D_\mu \Phi^\dagger)(\D^\mu \Phi)$,
\bea
	\l_\mathrm{int}^\mathrm{boson} & = & 
	g_\sf^2 \left( \Phi^\dagger T_\sf^a T_\sf^b \Phi \right) U_\mu^a {U^b}^\mu 
	+ g_\sg^2 \left( \Phi^\dagger T_\sg^a T_\sg^b \Phi \right) A_\mu^a {A^b}^\mu 
	\nonumber \\
	& & 
	+ g_\sf g_\sg \left[ \Phi^\dagger \left( T_\sf^a T_\sg^b + T_\sg^b T_\sf^a \right) \Phi \right] U^a_\mu {A^b}^\mu \; ,
        \label{eq:liboson}
\eea
The 4-vector ``current'' $J^a_\mu$ is,
\be
        {J^a_\sx}_\mu = \left\{
	\begin{array}{ll}
	{\displaystyle -i \left[ (\pd_\mu \Phi)^\dagger T_\sx^a \Phi - \Phi^\dagger T_\sx^a (\pd_\mu \Phi) \right]} & \mathrm{for \, boson} \\
	{\displaystyle \overline{\Psi} T_\sx^a \gamma_\mu \Psi} & \mathrm{for \, fermion} \\
	\end{array}
	\right. \, ,
        \label{eq:j}
\ee
with X : F, G and $\overline{\Psi} \equiv \Psi^\dagger \gamma_0$. 

Having the total lagrangian at hand, we can investigate further the dynamics of fluid and the contribution of  its interactions with another gauge fields. First of all, let us derive the equation of motion (EOM) of fluid that is our main interest in this Letter. It can be obtained from the lagrangian under consideration through the Euler-Lagrange equation in term of $U_\mu^a$, 
\be
	\frac{\pd \l}{\pd U_\nu^a} - \pd_\mu \, \frac{\pd \l}{\pd\left( \pd_\mu U_\nu^a \right) } = 0 \; ,
	\label{eq:ele}
\ee
Substituting Eq. (\ref{eq:l}) into Eq. (\ref{eq:ele}) yields the EOM,
\be
	\begin{array}{lcll}
	\pd^\nu S_{\mu\nu} & = & g_\sf {\J_\sf}_\mu & \mathrm{for \; Abelian} \; , \\
	\D^\nu S^a_{\mu\nu} & = & g_\sf {\J^a_\sf}_\mu & \mathrm{for \; non-Abelian} \; ,\\
	\end{array}
        \label{eq:eomf}
\ee
with the covariant current is,
\be
	{\J^a_\sf}_\mu \equiv \left\{
	\begin{array}{ll}
	{\displaystyle -i \left[ (\D_\mu \Phi)^\dagger T_\mathrm{F}^a \Phi - \Phi^\dagger T_\mathrm{F}^a (\D_\mu \Phi) \right]} & \mathrm{for \, boson} \\
	{\displaystyle {J^a_\sf}_\mu} & \mathrm{for \, fermion} \\
	\end{array}
	\right. \, .
        \label{eq:cc}
\ee
Moreover, the antisymmetric strength tensor in Eq. (\ref{eq:eomf}) implies that $\pd^\mu {\J^a_\sf}_\mu = 0$, that is ${\J^a_\sf}_\mu$ is a conserved current. 

Following the previous works in \cite{mahajan1, mahajan2, mahajan3}, we can take  fluid field to have a  relativistic form as follow,
\be
	U_\mu^a = (U^a_0, \U^a) \equiv u^a_\mu \phi \, \, \, \, \mathrm{with} \, \,\, \,  
	u_\mu \equiv \gamma^a (1, -\v^a) \; ,
	\label{eq:u}
\ee
where $u_\mu$ is  relativistic velocity and the relativistic factor $\gamma \equiv {(1 - |\v|^2)}^{{-1}/2}$, while $\v$ is the spatial velocity. On the other hand $\phi$ is a dimension one auxiliary field to keep correct dimension and should represent the fluid distribution in the system. We should remark that the index $a$ in $\gamma^a \v^a$ is only to label each flow field and not summing up all of them. Therefore, in the present model the fluid is described in terms of its kinematics and distribution function separately. It captures a complicated dynamics modeled as a fluid (like) system. In the case of non-Abelian fluid, we consider several flow fields labeled by the index $a$. We should remark that, $u_\mu$ in Eq. (\ref{eq:u}) differs slightly with the conventional notation. The reason of taking negative sign in the spatial component of $u_\mu$ will be clarified soon. However, more importantly the form still keeps Lorentz invariance.

Now we are ready to investigate in detail the EOM for general non-Abelian case in Eq. (\ref{eq:eomf}). For the sake of simplicity and concerning our current interest, Eq. (\ref{eq:eomf}) is rewritten in terms of double derivative, covariant current and the rest denoted by ``total force'' $F^a_\mu = (F^a_0, \F^a) = F^a_\mu(U,A)$. Hence summing up all components we simply have,
\be
	\pd^0 \left( \pd_\mu U^a_0 - \pd_0 U^a_\mu \right) - \pd^i \left( \pd_\mu U^a_i - \pd_i U^a_\mu \right) = g_\sf \left( {\J^a_\sf}_\mu + F^a_\mu \right) \; ,
	\label{eq:eomsu}
\ee
where, 
\bea
	F^a_\mu & \equiv & 
	f_\sf^{abc} \left[ \pd^0 \left( U_\mu^b U_0^c \right) - \pd^i \left( U_\mu^b U_i^c \right)\right] 
	\nonumber \\
	&& 
	- i \left( T_\sf^d {U^d}^0 + \frac{g_\sg}{g_\sf} T_\sg^d {A^d}^0 \right)
	\left( \pd_\mu U_0^a - \pd_0 U_\mu^a + g_\sf f_\sf^{abc} U_\mu^b U_0^c \right)
	\nonumber \\
	&& 
	+ i \left( T_\sf^d {U^d}^i + \frac{g_\sg}{g_\sf} T_\sg^d {A^d}^i \right)
	\left( \pd_\mu U_i^a - \pd_i U_\mu^a + g_\sf f_\sf^{abc} U_\mu^b U_i^c \right) \; ,
	\label{eq:eomfs}
\eea
using $T^aT^b = 2 i f^{abc} T^c$ in the case of SU($n$) that is our interest in further discussion. Greek (Latin) indices run over $0,1,2,3$ ($1,2,3$). Integrating out both sides in Eq. (\ref{eq:eomsu}) over $t$ ($x_i$) for $\mu=i$ ($\mu=0$), and writing $\J_\mu \equiv (\J_0, \Jb)$, we then have, 
\be
	\frac{\pd}{\pd t} \U^a - \grad U^a_0 = 
	-g_\sf \oint \d t \, \left( \Jb^a_\sf + \F^a \right) 
	= -g_\sf \oint \d \x \, \left( {\J^a_\sf}_0 + F^a_0 \right) \; ,
	\label{eq:eomb}
\ee
which is satisfying the condition for irrotational fluid, \ie the ``vorticity`` $\o$, 
\be
	\o^a \equiv \grad \times \U^a = 0 \; .
	\label{eq:irrot}
\ee
This condition is obtained as a non-trivial solution ($\U^a \ne 0$) from the terms $\pd^i \left( \pd_j U^a_i - \pd_i U^a_j \right) = \left[ \grad \left( \grad \cdot \U^a \right) - \grad^2 \U^a \right]_j = \left[ \grad \times \left( \grad \times \U^a \right) \right]_j $ in Eq. (\ref{eq:eomsu}), and $\pd_j U_i^a - \pd_i U_j^a = \epsilon_{jik} \left( \grad \times \U^a \right)^k$ in Eq. (\ref{eq:eomfs}) for $\mu = j \ne i$, while it is vanishing for $\mu = i$. Combining Eqs. (\ref{eq:u}) and (\ref{eq:eomb}),
\be
	\frac{\pd}{\pd t} \left( \gamma^a \v^a \phi \right) + \grad \left( \gamma^a \phi \right) = -g_\sf \oint \d \x \, \left( {\J^a_\sf}_0 + F^a_0 \right) \, \; .
	\label{eq:ree}
\ee
At non-relativistic limit, $\gamma \sim 1 + 1/2 |\v|^2$ and $\phi \sim 1$ for a constant $\phi$. These then yield,
\be
	\frac{\pd \v^a}{\pd t} + \frac{1}{2} \grad \left| \v^a \right|^2 = -g_\sf \left. \oint \d \x \, \left( {\J^a_\sf}_0 + F^a_0 \right) \right|_\mathrm{non-rel.} \; ,
	\label{eq:nree}
\ee
up to $O(|\v|^2)$ accuracy. Just to mention, one may also take a non-constant $\phi$, for instance the hot fluid which takes  $\phi_\mathrm{non-rel.}(T) \sim 1 + (5/2)(m/T)$ using large $T/m$ expansion \cite{mahajan1}, inducing some additional terms in both sides of Eq. (\ref{eq:nree}). Further, utilizing the vector identity $\frac{1}{2} \grad \left| \v \right|^2 = (\v \cdot \grad) \v + \v \times (\grad \times \v)$, finally we reach at,
\be
	\frac{\pd \v^a}{\pd t} + (\v^a \cdot \grad) \v^a = -g_\sf \left. \oint \d \x \, \left( {\J^a_\sf}_0 + F^a_0 \right)\right|_\mathrm{non-rel.} \; .
  	\label{eq:cee}
\ee
Obviously, Eq. (\ref{eq:cee}) reproduces the classical EOM for irrotational fluid. We argue that Eq. (\ref{eq:ree}) should be a general relativistic fluid equation. On the other hand, the ``current force'' $\J_\mu^a$ is induced by the existing matters surrounded by and interacting with the fluid, while $F_\mu^a$ is induced by the fluid self-interaction and the interacting gauge fields ($A^a_\mu$).  Therefore, the lagrangian in Eq. (\ref{eq:l}) with the fluid field $U^a_\mu$ having a form of Eq. (\ref{eq:u}) should describe a general relativistic fluid system interacting with another gauge fields and matters inside. Actually, the relative signs in the first and second terms in the left hand side of Eqs. (\ref{eq:ree}), (\ref{eq:nree}) and (\ref{eq:cee}) relies on the sign of $\U$ component in Eq. (\ref{eq:u}). This is the reason we should take the notation in Eq. (\ref{eq:u}) rather than the conventional one for relativistic velocity. 

To be more specific, let us consider a magnetofluid unification involving a non-Abelian SU(3) fluid contains particles of quarks and anti-quarks interacting with an electromagnetic field, \ie the lagrangian with SU(3)$_\sf$ $\otimes$ U(1)$_\sg$ gauge symmetry,
\bea
	\l & = & i \overline{\Q} \, \pds \Q - m_{\Q} \overline{\Q} \Q 
	- \frac{1}{4} S^a_{\mu\nu} {S^a}^{\mu\nu} 
	- \frac{1}{4} F_{\mu\nu} F^{\mu\nu} 
	+ g_\sf {J^a_\sf}_\mu {U^a}^\mu + q {J_\sg}_\mu A^\mu \; .
\eea
from Eqs. (\ref{eq:lmatter}), (\ref{eq:lfce}) and (\ref{eq:li}). Here $g_\sg$ is replaced with $q$ denoting the quark charge, $\Q$ represents the quark (color) triplet, ${J_\sf^a}^\mu = \overline{\Q} T_\sf^a \gamma^\mu \Q$ and $J_\sg^\mu = \overline{\Q} \gamma^\mu \Q$, while $T_\sf^a$'s belongs to the SU(3) Gell-Mann matrices. The model describes ``macroscopically'' non-Abelian fluid formed by dense gluon cloud surrounding the matters (quarks and anti-quarks) in an electromagnetic field. This fits the experimental clue from PHENIX Collaboration at the BNL-RHIC \cite{adeox}, far from being the weakly interacting collisionless plasma the deconfined quark gluon matter behaves more like a quark gluon fluid. This description is, however completely different with the hybrid magnetofluid model \cite{mahajan1,mahajan2} where the fluid flows denote the quarks and anti-quarks  interacting with the gluon fields. 

From Eq.  (\ref{eq:j}) the zero-th component of fermion current, \ie the current density, is ${\J_\sf}^a_0 = {J_\sf}^a_0 = \Q^\dagger T^a_\sf \Q \equiv \rho^a_\sf$. Using Eq. (\ref{eq:ree}) its relativistic dynamics obeys, 
\be
	\frac{\pd}{\pd t} \left( \gamma^a \v^a \phi \right) + \grad \left( \gamma^a \phi \right) = -g_\sf \oint \d \x \, \left( \rho^a_\sf + F^a_0 \right) \; .
	\label{eq:eomqgp}
\ee
Meanwhile $g_\sf$ should be associated with the fine structure constant of strong interaction $g_\sf^2 = 4\pi \, \alpha_s$ with $\alpha_s$ depends on the energy scale of physics under consideration \cite{pdb}. According to the experimental results \cite{shuryak}, the hot QGP is dense but seems to flow with tiny viscosity approximating the ideal fluid. This draws $\grad \cdot \U^a \sim 0$ affecting the term $f_\sf^{abc} \pd^i \left( U_0^b U_i^c \right)$ in Eq. (\ref{eq:eomfs}), and $\o^a \sim 0$ satisfying Eq. (\ref{eq:irrot}). Moreover, as seen in the above lagrangian the gluon fluid itself does not feel the electromagnetic force, but the quarks and anti-quarks do. In contrary, the electromagnetic force still contributes to the EOM in Eq.  (\ref{eq:eomqgp}), but it is suppressed by a factor of $q/{g_\sf} \propto \sqrt{\alpha/{\alpha_s}} \sim O({10}^{-1})$ since the quark charge is proportional to the electric charge $e = \sqrt{\alpha/{4\pi}}$. Therefore the electromagnetic force contributes about few percents, and might be neglected for a good accuracy.

For completeness, we can derive the total energy due to fluid and electromagnetic fields in the above case. The lagrangian density for the fluid and electromagnetic fields is, 
\be
	\l = 
	- \frac{1}{4} S^a_{\mu\nu} {S^a}^{\mu\nu} 
	- \frac{1}{4} F_{\mu\nu} F^{\mu\nu} 
	+ g_\sf {J^a_\sf}_\mu {U^a}^\mu + q {J_\sg}_\mu A^\mu 
\ee
We can define ${E^a_\sf}_i \equiv -S^a_{0i}$ and $\epsilon_{jik} {B^a_\sf}^k \equiv S^a_{ji}$ with $i \ne j$ denoting the electric- and magnetic-like fields of fluid analogous to the usual electric and magnetic fields $E_i = -F_{0i}$ and $\epsilon_{jik} B^k = F_{ji}$. Then, $S^a_{\mu\nu} {S^a}^{\mu\nu} = -2 \left( \left| \E^a_\sf \right|^2 - \left| \B^a_\sf \right|^2 \right)$ in analogy to $F_{\mu\nu} F^{\mu\nu} = -2 \left( \left| \E \right|^2 - \left| \B \right|^2 \right)$. Due to the condition in Eq. (\ref{eq:irrot}), $\B_\sf^a = \frac{1}{2} g_\sf f^{abc} \U^b \times \U^c$,  while $\E_\sf^a = -g_\sf \oint \d \x \, \left( \rho^a_\sf + F^a_0 \right)$ from the EOM in Eq. (\ref{eq:eomf}). Hence, omitting the small electromagnetic current term we can easily deduce the Hamiltonian density,
\bea
	\H & = &
	\frac{1}{2} g_\sf^2 \left( 
	\left| \oint \d \x \, \left( \rho^a_\sf + F^a_0 \right) \right|^2 
	+ \frac{1}{2} \left| f^{abc} \U^b \times \U^c \right|^2 \right)
	+ g_\sf \gamma^a \phi \left( \rho^a_\sf + \v^a \cdot \Jb^a_\sf \right)
	\nonumber \\
	& & 
	+ \frac{1}{2} \left( \left| \E \right|^2 + \left| \B \right|^2 \right) \; .
	\label{eq:h}
\eea
In the simplest case of pure gluon fluid, only the first and second terms will remain,
\be
	\H = \frac{1}{2} g_\sf^2 \left( 
	\left| \oint \d \x \, F^a_0 \right|^2 
	+ \frac{1}{2} \left| f^{abc} \U^b \times \U^c \right|^2 \right) \; ,
\ee
since ${\J^a_\sf}_\mu = 0$ in the absence of matter. Taking its non-relativistic limit, the total energy is simply $H = \int \d^3x \, \H_\mathrm{non-rel.}$. For example having particular temperature dependent $\phi$ would actually enable us to calculate any physical observables (free energy, entropy, etc) through the partition function density $\Z = \mathrm{Tr} \left[ \mathrm{exp}(\H/T) \right]$. 

We have studied the magnetofluid unification using lagrangian approach and imposing the gauge principle in Abelian and non-Abelian cases. The model provides new insights into macroscopic dynamics of relativistic hot fluids relevant for QGP. An example of SU(3)$_\sf \otimes$ U(1)$_\sg$ model, in particular, may describe some of the puzzles posed by the data from the RHIC indicating that there is a very rapid thermalization in the collisions, after which a fluid with very low viscosity and large transport coefficients seems to be produced, and surprisingly the produced medium seems to be strongly interacting \cite{adeox}. Recently, the ALICE experiment at the LHC is also challenging to resolve this puzzle \cite{alice}. It is also argued that the formalism leads to different physical consequences than the other ones derived from the hybrid magnetofluid model. However, all consequences here arise naturally from  first principle without any fine tuning. The results should lead to some interesting phenomenon either in QGP or fluid based cosmology. Such studies, including simulations based on numerical computation for particular cases, are in progress. 

We greatly appreciate fruitful discussion with N. Riveli, H.B. Hartanto and A. Oxalion throughout the work.  This work is partially funded by the Indonesia Ministry of Research and Technology and the Riset Kompetitif LIPI in fiscal year 2007 under Contract no.  11.04/SK/KPPI/II/2007.

\end{document}